\documentclass[aps,twocolumn,superscriptaddress,nofootinbib]{revtex4-2}
\usepackage{graphicx}
\usepackage{dcolumn}
\usepackage{bm}
\usepackage[utf8]{inputenc}
\usepackage{float}
\usepackage{hyperref}
\usepackage{amsmath, amsthm, amsfonts,amssymb}
\usepackage[svgnames]{xcolor}
\usepackage{mathtools}
\usepackage{caption}
\usepackage{subcaption}
\usepackage{xcolor}
\usepackage{soul}
\usepackage{xcolor}
\usepackage{appendix}
\usepackage{orcidlink}
\usepackage{booktabs}
\usepackage[normalem]{ulem}
\usepackage{comment}

\definecolor{teal}{RGB}{0, 128, 128}
\hypersetup{ colorlinks=true, linkcolor=teal, citecolor=teal, urlcolor=teal}

\begin{document}

\title{Vacuum polarization in the Schwarzschild black hole with a global monopole}

\author{Leonardo G. Barbosa \orcidlink{0009-0007-3468-3718}}
\email{leonardo.barbosa@posgrad.ufsc.br}
\affiliation{Departamento de Física, CFM - Universidade Federal de \\ Santa Catarina; C.P. 476, CEP 88.040-900, Florianópolis, SC, Brazil}

\author{Victor Hugo M. Ramos \orcidlink{0000-0002-1763-8663}}
\email{vhmarques@usp.br}
\affiliation{Instituto de Física, Universidade de São Paulo, Rua do Matão 1371,
São Paulo, 05508-090, Brasil}

\author{João Paulo M. Pitelli \orcidlink{0000-0002-5728-5014}}
\email{pitelli@unicamp.br}
\affiliation{Departamento de Matemática Aplicada, Universidade Estadual de \\ Campinas, 13083-859 Campinas, São Paulo, Brazil}

\begin{abstract}
We investigate vacuum polarization on the event horizon of a Schwarzschild black hole carrying a global monopole. For a massless scalar field $\Psi$ in the Hartle-Hawking state and with arbitrary curvature coupling, we compute the renormalized vacuum expectation value $\langle \Psi^2 \rangle_{\textrm{ren}}$. The monopole produces a solid-angle deficit and makes the spacetime non-Ricci-flat. Working perturbatively in the monopole parameter $\eta$ and retaining terms through $O(\eta^2)$, we find that $\langle \Psi^2 \rangle_{\textrm{ren}}$ on the horizon splits into two contributions: a genuinely monopole-induced term evaluated at the horizon and the usual Schwarzschild result--with the event horizon radius modified by the presence of $\eta$.  We also investigate whether an analogous decomposition holds for $\langle T^{\mu}_{\phantom{\mu}\mu}\rangle_\textrm{ren}$ when it is determined by this method. Our result parallels earlier analyses of Schwarzschild black holes pierced by a cosmic string.
\end{abstract}

 \maketitle

\section{Introduction}\label{Introduction}

Global monopole spacetimes arise naturally in grand unified theories (GUTs) as a consequence of the spontaneous breaking of a global $O(3)$ symmetry in a triplet scalar field $\phi^a$ ($a=1,2,3$) during a phase transition in the early Universe~\cite{vilenkin}. The resulting geometry exhibits a solid-angle deficit: spheres of radius $r$ have area $4\pi(1-\eta^2)\,r^2$, so that the missing solid angle is $4\pi\eta^2$, where $0<\eta<1$ is determined by the symmetry-breaking scale (for monopoles formed at the grand unified scale one typically expects $\eta^2\sim 10^{-5}$~\cite{ross1984grand}). Unlike the spacetime of a straight cosmic string--which features an azimuthal deficit but remains locally flat outside its core--the global monopole spacetime is genuinely curved and is supported by a nonvanishing energy density that falls off as $\sim r^{-2}$. In an idealized description, the metric can be written as
\begin{equation}
ds^2 = -dt^2 + dr^2 + (1-\eta^2)\,r^2 d\Omega^2,
\label{metric1}
\end{equation}
where $d\Omega^2$ denotes the line element on the unit round 2-sphere. In this form, the solid-angle deficit is explicit. An equivalent parametrization, frequently employed in the literature, is obtained by introducing the rescaled coordinates
\begin{equation}
t \to  \sqrt{1-\eta^2}\,t,
\qquad
r\to  \frac{r}{\sqrt{1-\eta^2}},
\label{change of coordinates}
\end{equation}
in terms of which the line element becomes\footnote{Both coordinate parametrizations are used in the literature. Throughout most of this paper we work in coordinates that reduce to Eq.~\eqref{metric2} in the limit $M\to 0$ (with $M$ the black hole mass), which allows us to use intermediate expressions from Ref.~\cite{mazzitelli}. When needed, we switch to coordinates that reduce to Eq.~\eqref{metric1}, as this makes the comparison with Candelas' Schwarzschild result~\cite{candelas} more transparent.}

\begin{equation}
ds^2 = -(1-\eta^2)dt^{2} + \frac{dr^{2}}{(1-\eta^2)} + r^{2} d\Omega^2.
\label{metric2}
\end{equation}

The spacetime also contains a naked timelike singularity at $r=0$. As a consequence, the renormalized vacuum polarization $\langle \Psi^2\rangle_{\textrm{ren}}$ for a quantum scalar field is sensitive not only to the nontrivial topology and curved geometry of the monopole background~\cite{mazzitelli}, but also to the singular structure at the origin, as encoded in the boundary condition imposed on the field at $r=0$~\cite{pitelli1,pitelli2}.

In Ref.~\cite{mazzitelli}, Mazzitelli and Lousto calculated the renormalized expectation value $\langle \Psi^2\rangle_{\textrm{ren}}$  for a massless scalar field with arbitrary curvature coupling, imposing Dirichlet boundary conditions at the origin. It was shown that the vacuum fluctuations diverge as one approaches the singularity at $r=0$ through the form
\begin{equation}
\label{eq:MLterm}
\left\langle \Psi^{2}\right\rangle_{\mathrm{ren}}^{\mathrm{ML}}
=
-\frac{\eta^{2}\!\left(p-2\xi q\right)}{8\sqrt{2}\pi^{2}r^{2}}
-\left(\xi-\frac{1}{6}\right)\frac{\eta^{2}}{8\pi^{2}r^{2}}\ln\!\left(\mu^{2}r^{2}\right),
\end{equation}
up to second order in $\eta$. In Ref.~\cite{pitelli1}, the same observable was computed for a general boundary condition at the origin, and it was shown that non-Dirichlet choices generate an additional contribution to $\left\langle \Psi^{2}\right\rangle_{\mathrm{ren}}$ already at $O(\eta^{0})$. In Ref.~\cite{pitelli2}, in turn, the divergence of the transition rate of an Unruh-DeWitt detector held at a fixed radius $r_0$ was related to the behavior of $\langle \Psi^2 \rangle_{\textrm{ren}}$ (for an arbitrary choice of boundary condition) in the limit $r_0 \to 0$. It was shown that the transition rate and the vacuum fluctuations are amplified by the singularity in the same manner.

Taken together, these results show that the renormalized vacuum polarization $\langle \Psi^2\rangle$ provides a convenient and physically informative probe of quantum effects in global monopole spacetimes. Moreover, it captures essential qualitative features of the renormalized stress-energy tensor $\langle T_{\mu\nu} \rangle$~\cite{winstanley} while being technically simpler to evaluate, making $\langle \Psi^2\rangle$ a natural observable in global monopole geometries and their generalizations.

A particularly relevant generalization is the Schwarzschild-global monopole spacetime. Its construction closely parallels that of the Schwarzschild solution~\cite{barriola,dadhich,zloshchastiev}: one assumes a static, spherically symmetric geometry and imposes
\begin{equation}
R^{0}{}_{0} = R^{1}{}_{1} = 0,
\qquad
R_{01}=0,
\end{equation}
while allowing $R^{2}{}_{2}\neq 0$. Via Einstein's equations, this corresponds to an effective energy-momentum tensor with components 
\begin{equation}
T^{0}{}_{0} = T^{1}{}_{1} = - \frac{\eta^2}{8\pi r^2},
\qquad
T^{2}{}_{2} = T^{3}{}_{3} = 0,
\end{equation}
and leads to the line element~\cite{barriola}

\begin{equation}
  ds^{2}=-\left(1-\eta^{2}-\frac{2M}{r}\right)dt^{2}+\frac{dr^{2}}{\left(1-\eta^{2}-\frac{2M}{r}\right)}+r^{2}d\Omega^{2},
\label{SMBH}
\end{equation}
where $M$ is a parameter related to the black hole mass.

A natural question is how physical observables, such as the renormalized vacuum polarization $\langle \Psi^2\rangle_{\textrm{ren}}$ and the renormalized stress-energy tensor $\langle T_{\mu\nu}\rangle_{\textrm{ren}}$, respond when the nontrivial topology associated with a global monopole is combined with the presence of an event horizon. For instance, Ref.~\cite{dadhich} shows that, for the Schwarzschild-global monopole metric, the Hawking temperature is scaled by a factor $(1-\eta^2)^2$ (for fixed mass $M$). This illustrates how the interplay between a deficit-angle geometry and an event horizon impacts physical observables.

Candelas, in Ref.~\cite{candelas}, calculated $\langle \Psi^2\rangle_{\textrm{ren}}$ for a massless, conformally coupled scalar field on the event horizon of a Schwarzschild black hole. Exploiting that in the Hartle-Hawking vacuum (corresponding to a black hole of mass $M$ in thermal equilibrium with a bath of blackbody radiation) the Euclidean Green's function admits an expansion in Legendre functions, Candelas obtained the near-horizon vacuum polarization,
\begin{equation}
\langle \Psi^2\rangle_{\textrm{ren}}(r\to 2M)=\frac{1}{192\pi^2 M^2}.
\end{equation}

In Refs~\cite{ottewill1,ottewill2}, Ottewill and Taylor generalized this result to a Schwarzschild black hole threaded by a
cosmic string. They found that, for a minimally coupled scalar field, 
\begin{equation}\label{eq:Psi2Ottewill}
\langle \Psi^2\rangle_{\textrm{ren}}^\textrm{horizon}=\frac{1}{192\pi^2M^2}\left(1-\frac{1-\alpha^2}{\alpha^2\cos^2{\theta}}\right).
\end{equation}
This result corresponds to the sum of Candelas' result with the cosmic string contribution~\cite{grats}  for a fixed $\theta$ on the horizon. Additionally, we emphasize that this angular dependence is expected given the presence of the cosmic string, which breaks the spherical symmetry.

In this paper, we compute \(\langle \Psi^2\rangle_{\textrm{ren}}^{\textrm{horizon}}\) for a Schwarzschild black hole carrying a global monopole. Following the approach of Candelas, we find that the conclusion of Refs.~\cite{ottewill1,ottewill2} still holds: the vacuum polarization can be decomposed into two independent contributions, one inherited from the pure monopole spacetime and evaluated at the horizon, and the other corresponding to the pure Schwarzschild spacetime. As a first step toward the analysis of the renormalized stress-energy tensor, we also investigate whether this additive decomposition is reflected in the conformal anomaly, postponing a full treatment of the renormalized stress-energy tensor for future work.

\section{Green's function}
In this section we study the Green's function $G(x,x^{\prime})$ for a massless scalar field $\Psi$ with nonminimal coupling in the spacetime of a global monopole Schwarzschild black hole  given by the metric~(\ref{SMBH}). We then consider the Klein-Gordon equation in its covariant form with nonminimal coupling
\begin{equation}
  \left(\nabla_{\mu}\nabla^{\mu}-\xi R\right)G\left(x,x^{\prime}\right)=-\frac{1}{\sqrt{g}}\delta^{\left(4\right)}\left(x-x^{\prime}\right)
  \label{KG}
\end{equation}

We are interested in the Hartle-Hawking vacuum. In this context, we consider the Wick rotation $t\rightarrow i\tau$ in the line element ~(\ref{SMBH}). Seeking to eliminate conical singularities, we are led to take $\tau$ with a period $\beta=2\pi/ \kappa$, where $\kappa=f^{\prime}(r_h)/2=\frac{(1-\eta^2)^2}{4M}$ is the surface gravity of the black hole, with $r_h=2M/(1-\eta^2)$ defining the event horizon. In this case, we can adopt the following ansatz for the Euclidean Green’s function:
\begin{equation}
\begin{aligned}
    G(x,x') &= \frac{1}{\beta}\sum_{n=-\infty}^{\infty}
    e^{i\omega_n(\tau-\tau')} \\
    &\times
    \sum_{\ell m} 
    Y_{\ell m}(\theta,\phi)\,
    Y_{\ell m}^{*}(\theta',\phi')\,
    g_{n\ell}(r,r')
    \label{pregreen}
\end{aligned}
\end{equation}
where $\omega_{n}=2\pi n /\beta$, with $n\in\mathbb{Z}$. Substituting Eq.~(\ref{pregreen}) into Eq.~(\ref{KG}), we arrive at
\begin{equation}
\begin{aligned}
   &\left\{ \frac{d}{dr}\left(\Delta\left(r\right)\frac{d}{dr}\right)-\frac{n^{2}\kappa^{2}r^{4}}{\Delta\left(r\right)}-\ell\left(\ell+1\right)-2\xi\eta^{2}\right\} g_{n\ell}\left(r,r^{\prime}\right)\\&=-\delta\left(r-r^{\prime}\right),
\end{aligned}
\end{equation}
in which we have defined the following auxiliary quantity: $\Delta\left(r\right)=\left(1-\eta^{2}\right)r\left(r-r_{\text{h}}\right)$.

It is convenient to write the radial equation in terms of a new radial variable \cite{ottewill1}
\begin{equation}
\rho= \frac{2r}{r_{h}}-1,
\label{rho}
\end{equation} where now the event horizon is located at $\rho=1$. Under this change of variable, we can rewrite the radial equation as
\begin{equation}
\begin{aligned}
 &\left\{ \frac{d}{d\rho}\left(\left(\rho^{2}-1\right)\frac{d}{d\rho}\right)-\lambda_{\ell,\xi,\eta}\left(\lambda_{\ell,\xi,\eta}+1\right)-\frac{n^{2}\left(1+\rho\right)^{4}}{16\left(\rho^{2}-1\right)}\right\} \\&\times g_{n\ell}\left(\rho,\rho^{\prime}\right)=-\frac{1}{M}\delta\left(\rho-\rho^{\prime}\right),
 \label{eq.radial}
\end{aligned}
\end{equation}
where the parameter $\lambda_{\ell,\xi,\eta}$ is an effective angular momentum quantum number that incorporates the dependence on both the coupling constant $\xi$ and the global monopole parameter $\eta$. This quantity is implicitly determined by the relation
\begin{equation}
    \lambda_{\ell,\xi,\eta}\left(\lambda_{\ell,\xi,\eta}+1\right)
    = \frac{\ell(\ell+1) + 2\xi\eta^{2}}{(1 - \eta^{2})},
\end{equation}
which generalizes the standard angular momentum term, $\ell(\ell +1)$, to account for the effects of the global monopole and the nonminimal coupling represented by $\xi$.

For $n = 0$, the two solutions of the homogeneous equation are the Legendre functions of the first and second kind, respectively $P_{\lambda_{\ell,\xi,\eta}}\left(\rho\right)$ and $Q_{\lambda_{\ell,\xi,\eta}}\left(\rho\right)$. Hence, the solution for Eq.(\ref{eq.radial}) reads 
\begin{equation}
g_{0\ell}(\rho,\rho') =
\begin{cases}
A_{<}P_{\lambda_{\ell,\xi,\eta}}(\rho)+B_{<}Q_{\lambda_{\ell,\xi,\eta}}(\rho), & \rho<\rho',\\[6pt]
A_{>}P_{\lambda_{\ell,\xi,\eta}}(\rho)+B_{>}Q_{\lambda_{\ell,\xi,\eta}}(\rho), & \rho>\rho',
\end{cases}
\end{equation}
where $A_{<}$, $B_{<}$, $A_{>}$, and $B_{>}$ are constants to be determined. Indeed, imposing regularity of $g_{0\ell}(\rho,\rho^{\prime})$ near the horizon and the infinity fixes $A_{>}=B_{<}=0$. Additionally, this solution must be continuous at $\rho=\rho^{\prime}$ and satisfy the jump condition for its first derivative, \begin{align}
\lim_{\delta\rightarrow0}\left[\left(\rho^{2}-1\right)\frac{dg_{0\ell}}{d\rho}\right]_{\rho^{\prime}-\delta}^{\rho^{\prime}+\delta}=-\frac{1}{M}.
\end{align} These requirements completely characterize the solution, which is then given by
\begin{equation}
    g_{0\ell}\left(\rho,\rho^{\prime}\right)=\frac{1}{M}P_{\lambda_{\ell,\xi,\eta}}\left(\rho_{<}\right)Q_{\lambda_{\ell,\xi,\eta}}\left(\rho_{>}\right),
    \label{GreenI}
\end{equation}
where we have denoted $\rho_{>} = \max\{\rho,\rho^{\prime}\}$ and $\rho_{<} = \min\{\rho,\rho^{\prime}\}$.

For $n\neq 0$, we can perform a treatment for the homogeneous equation using the Frobenius method. In the asymptotic case near the event horizon, we can write 
\begin{subequations}
\begin{align}
p_{\lambda_{\ell,\xi,\eta}}^{\lvert n\rvert}(\rho) &\sim (\rho-1)^{\frac{\lvert n\rvert}{2}}, \qquad \rho\to 1,\\
q_{\lambda_{\ell,\xi,\eta}}^{\lvert n\rvert}(\rho) &\sim (\rho-1)^{-\frac{\lvert n\rvert}{2}}, \qquad \rho\to 1.
\end{align}
\end{subequations}
Similarly to the previous case, after imposing regularity, continuity, and the jump condition, we obtain
\begin{equation}
    g_{n\ell}\left(\rho,\rho^{'}\right)=\frac{1}{2\left|n\right|M}p_{\lambda_{\ell,\xi,\eta}}^{\left|n\right|}\left(\rho_{<}\right)q_{\lambda_{\ell,\xi,\eta}}^{\left|n\right|}\left(\rho_{>}\right). 
\end{equation}

In summary, the complete radial Green's function is given by
\begin{equation}
g_{n\ell}(\rho,\rho') =
\begin{cases}
\dfrac{1}{M}\,P_{\lambda_{\ell,\xi,\eta}}(\rho_{<})\,Q_{\lambda_{\ell,\xi,\eta}}(\rho_{>}), & n=0,\\[8pt]
\dfrac{1}{2\lvert n\rvert M}\,p_{\lambda_{\ell,\xi,\eta}}^{\lvert n\rvert}(\rho_{<})\,q_{\lambda_{\ell,\xi,\eta}}^{\lvert n\rvert}(\rho_{>}), & n\neq 0.
\end{cases}
\end{equation}

Since our goal is to evaluate the relevant quantities at the event horizon, it should be noted that only the terms corresponding to $n=0$ contribute in this limit. For the region $\rho > \rho'$, the radial Green’s function satisfies (\ref{GreenI}). By taking the limit $\rho' \rightarrow 1$ and recalling that $P_{\lambda_{\ell,\xi,\eta}}\left(1\right)=1$, the expression for the Green’s function simplifies considerably.  Regarding the angular sector of the Green's function, we consider the spherical harmonic addition theorem given by
\begin{equation}
    \sum_{m=-\ell}^{\ell}Y_{\ell m}\left(\theta,\phi\right)
    Y_{\ell m}^{*}\left(\theta',\phi'\right)
    =\frac{2\ell+1}{4\pi}P_{\ell}\left(\cos\left(\gamma\right)\right),
\end{equation}
where the angle $\gamma$ represents the  angular separation of points and is defined by
\begin{equation}
\cos\left(\gamma\right)
=\cos\left(\theta\right)\cos\left(\theta'\right)
+\sin\left(\theta\right)\sin\left(\theta'\right)
\cos\left(\phi-\phi'\right).
\end{equation}
This relation allows the angular dependence of the Green’s function to be expressed in terms of Legendre polynomials.

For $n=0$ we have $\omega_{0}=0$ in Eq.~(\ref{pregreen}) and the Green’s function can then be written in the compact form,
\begin{equation}
G\left(x,x'\right)=\frac{\left(1-\eta^{2}\right)^{2}}{32\pi^{2}M^{2}}
\sum_{\ell=0}^{\infty}\left(2\ell+1\right)
P_{\ell}\left(\cos\left(\gamma\right)\right)
Q_{\lambda_{\ell,\xi,\eta}}\left(\rho\right).
\end{equation}
In the coincidence limit, where $\theta=\theta'$ and $\phi=\phi'$, the angular separation vanishes, leading to $P_{\ell}\left(1\right)=1$. In this case, the Green’s function simplifies to a purely radial expression,
\begin{equation}
G\left(\rho\right)=\frac{\left(1-\eta^{2}\right)^{2}}{32\pi^{2}M^{2}}
\sum_{\ell=0}^{\infty}\left(2\ell+1\right)
Q_{\lambda_{\ell,\xi,\eta}}\left(\rho\right).
\end{equation}
This form is particularly convenient for analyzing the behavior of the field near the horizon and for evaluating regularized quantities.

To facilitate a direct comparison with the formalism adopted by Mazzitelli and Lousto~\cite{mazzitelli}, we introduce the parameter 
$\lambda_{\ell,\xi,\eta}=\nu_{\ell,\xi,\eta}-1/2$. Furthermore, we introduce the integral representation for the Legendre functions,
\begin{align}\label{eq:LegendreRepresentation}
    Q_{\nu_{\ell,\xi,\eta}-\frac{1}{2}}\left(\cosh\left(\chi\right)\right)=\frac{1}{\sqrt{2}}\int_{\chi}^{\infty}dt\frac{e^{-\nu_{\ell,\xi,\eta}t}}{\sqrt{\cosh\left(t\right)-\cosh\left(\chi\right)}},
\end{align}
which renders the Green's function to the form 
\begin{multline}
   G\left(\rho\right)= \frac{\left(1-\eta^{2}\right)^{2}}{32\pi^{2}M^{2}}\frac{1}{\sqrt{2}}\int_{\chi}^{\infty}\frac{dt}{\sqrt{\cosh\left(t\right)-\cosh\left(\chi\right)}}\\ \times \sum_{\ell=0}^{\infty}\left(2\ell+1\right)e^{-\nu_{\ell,\xi,\eta}t},
   \label{eq:GreensFunction}
\end{multline} 
where $\rho = \cosh{\chi}$. This form is particularly useful when considering, for example, the GUT-scale scenario of very small curvature $\eta^2 \ll 1$. In this case the spectral parameter $\nu_{\ell,\xi,\eta}$ simplifies to\footnote{Perturbatively expanding $\nu_{\ell,\xi,\eta}$ to $O(\eta^2)$ makes the exponent linear in $\ell$ up to that order, thereby enabling the sum over $\ell$ to be evaluated as a geometric series. The expansion is justified for \(\eta^2 \ll 1\), which holds, for example, for GUT-scale monopoles~\cite{barriola}. Moreover, we expect our approximate expressions to remain valid even for values as large as \(10^{-2}\) or possibly \(10^{-1}\), as was shown in the special case \(\xi=1/8\) for a global monopole spacetime in Ref.~\cite{mazzitelli}.}
\begin{align}
    \nu_{\ell,\xi,\eta}\approx\left(\ell+\frac{1}{2}\right)\left(1+\frac{\eta^{2}}{2}\right)+\frac{\left(2\xi-\frac{1}{4}\right)\eta^{2}}{2\ell+1}
\end{align} and we can use the geometric series on the exponential contribution to  to rewrite the Green's function as 
\begin{widetext}
    \begin{align}
G\left(\rho\right)\approx\frac{1}{32\pi^{2}M^{2}}\frac{1}{\sqrt{2}}\int_{\chi}^{\infty}\frac{e^{-t/2}}{\sqrt{\cosh\left(t\right)-\cosh\left(\chi\right)}}\frac{1+e^{-t}}{\left(1-e^{-t}\right)^{2}}\left\{ 1-2\eta^{2}\left(1+\frac{t}{1-e^{-2t}}\left[\xi\left(1-e^{-t}\right)^{2}+e^{-t}\right]\right)\right\} dt. 
\label{GreenE}
    \end{align}
\end{widetext} This expression represents the Green’s function evaluated  close to the event horizon for a massless, nonminimally coupled scalar field in the background of a black hole with a global monopole in the approximation of $\eta^2 \ll 1$. It explicitly depends on the black hole mass parameter $M$ and on the monopole parameter $\eta$, reflecting how both quantities influence the near-horizon behavior of the scalar field.

\section{Vacuum polarization}
Now that we have obtained the full Green's function near the horizon, we can perform the computation of the renormalized vacuum fluctuation of the field squared. We recall that this quantity is defined as the coincident-point limit of the renormalized Green's function,   
\begin{align}
    \langle \Psi^2 \rangle_{\text{ren}} = \lim_{x^{\prime}\to x} G_{\text{ren}}(x,x^{\prime}),
\end{align} 
where the renormalized Green's function is given by $G_{ren}(x,x^{\prime}) = G(x,x^{\prime}) - G_{\text{sing}}(x,x^{\prime})$, with $G(x,x^{\prime})$ being the Green's function associated with the Klein-Gordon Eq.(\ref{KG}) and $G_{\text{sing}}(x,x^{\prime})$ being defined by~\cite{mazzitelli}
\begin{multline}
G_{\text{sing}}\left(x,x'\right)=\frac{1}{16\pi^{2}}\left[\frac{2}{\sigma\left(x,x'\right)}\right.\\ \left.+\left(\xi-\frac{1}{6}\right)R\ln\left(\frac{1}{2}\mu^{2}\sigma\left(x,x'\right)\right)\right].
\end{multline}
Here, $\sigma\left(x,x'\right)$ is half the square of the geodesic distance between $x$ and $x^{\prime}$, explicitly $\sigma(x,x^{\prime})=s^{2}(x,x^{\prime})/2$, and $\mu$ is an arbitrary renormalization scale. After subtracting all the ultraviolet behavior present in the Green's function, $\langle \Psi^2 \rangle_{\text{ren}}$ becomes a finite quantity. Therefore, we start by addressing the subtraction of the singularities in (\ref{eq:GreensFunction}).

Considering the near-horizon region, we compute the radial geodesic distance between $r_{h}$ and $r_{h} + \varepsilon$ in the first order of $\varepsilon$ \cite{Poisson_2011,ottewill1}, 
\begin{equation}
  s=\int_{r_{h}}^{r_{h}+\varepsilon}\frac{dr^{\prime}}{\sqrt{f(r^{\prime})}}=\frac{(2M\varepsilon)^{\frac{1}{2}}}{(1-\eta^{2})}\left[2+\frac{(1-\eta^{2})\varepsilon}{6M}+\mathcal{O}(\varepsilon^{2})\right],
\end{equation} where it is useful to notice that by Eq.(\ref{rho}) we can express $\varepsilon=r_{h}\left(\rho-1\right)/2$. As we are interested in studying the behavior of observables near the horizon and in the approximation of $\eta^2 \ll 1$, we consider the expression above in the leading order of $(\rho-1)$ and $\eta^2$. At this regime, the first contribution to the singular Green's function reads
\begin{align}\label{eq:InverseSigma1}
\dfrac{1}{8\pi^2}\dfrac{1}{\sigma(\rho)} \approx -\frac{\left(1-3\eta^{2}\right)}{192\pi^{2}M^{2}}+\frac{\left(1-3\eta^{2}\right)}{32\pi^{2}M^{2}(\rho-1)}.
\end{align} To facilitate the subtraction and the coincidence limit procedure, we recast the above expression to a form that resembles the Green's function (\ref{GreenE}) by considering the series representation $(\rho-1)^{-1}=\sum_{\ell=0}^{\infty}\left(2\ell+1\right)Q_{\ell}\left(\rho\right)$, and the integral representation (\ref{eq:LegendreRepresentation}). As a result, we obtain
\begin{multline}
\frac{1}{8\pi^{2}\sigma}\approx -\frac{(1-3\eta^{2})}{192\pi^{2}M^{2}}\\
+\frac{(1-3\eta^{2})}{32\pi^{2}M^{2}}\frac{1}{\sqrt{2}}
\int_{\chi}^{\infty}\frac{t\,e^{-t/2}}{\sqrt{\cosh t-\cosh\chi}}
\frac{1+e^{-t}}{(1-e^{-t})^{2}}\,dt.
\label{G_sing_I}
\end{multline}

We now evaluate the logarithmic contribution to the Schwinger-DeWitt expansion, defined as 
\begin{align}
    G_{\log}(x,x') = \frac{R}{16\pi^{2}}\left(\xi-\frac{1}{6}\right)\ln\left(\frac{1}{2}\mu^{2}\sigma\left(x,x^{\prime}\right)\right).
\end{align} In fact, by replacing the curvature scalar and consistently managing the leading-order regime in both $(\rho-1)$ and $\eta^2$ parameters, we directly obtain
\begin{align}\label{eq:GLog}
    G_{\log}(\rho) \approx \frac{\left(\xi-\frac{1}{6}\right)\eta^{2}}{8\pi^{2}M^{2}\left(\rho+1\right)^{2}}\ln\left(2\mu^{2}M^{2}(\rho-1)\right).
\end{align}
A more suitable form of expressing Eq.~(\ref{eq:GLog}) is achieved by rewriting it in terms of the Legendre function $Q_0$. Indeed, by the logarithmic representation 
\begin{align}
Q_{0}\left(\cosh\chi\right)=\frac{1}{2}\ln\left(\frac{\cosh\chi+1}{\cosh\chi-1}\right),
\end{align} 
and the integral representation in Eq.~(\ref{eq:LegendreRepresentation}), we arrive at 
\begin{multline}
    G_{\log}(\rho) \approx \left(\xi-\frac{1}{6}\right)\frac{\eta^{2}\ln\left(2\mu^{2}M^{2}\left(\rho+1\right)\right)}{8\pi^{2}M^{2}\left(\rho+1\right)^{2}} \\ -\frac{\left(\xi-\frac{1}{6}\right)\eta^{2}}{4\pi^{2}M^{2}\left(\rho+1\right)^{2}}\frac{1}{\sqrt{2}}\int_{\chi}^{\infty}dt\frac{e^{-t/2}}{\sqrt{\cosh\left(t\right)-\cosh\left(\chi\right)}}.
    \label{G_sing_log}
\end{multline}

Finally, we have that (\ref{G_sing_I}) and (\ref{G_sing_log}) completely characterize the singular structure of the Green's function and we obtain, by direct calculation, the value of the horizon.
\begin{widetext}
    \begin{equation}
        \left\langle \Psi^{2}\right\rangle _{\text{ren}}\approx -\frac{\eta^{2}\left(p-2\xi q\right)}{32\sqrt{2}\pi^{2}M^{2}}-\frac{\left(\xi-\frac{1}{6}\right)\eta^{2}}{32\pi^{2}M^{2}}\ln\left(4\mu^{2}M^{2}\right) +\frac{1-3\eta^{2}}{192\pi^{2}M^{2}},
        \label{eq:phi2_final}
    \end{equation}
\end{widetext}
where the constants $p$ and $q$ are given by
\begin{subequations}
\begin{align}
p&=\int_{0}^{\infty}\frac{dte^{-t/2}}{\sqrt{\cosh t-1}}
\left[\frac{1}{3}+\left(\frac{t}{\sinh t}-1\right)\frac{1+e^{-t}}{(1-e^{-t})^{2}}\right], \\
q&=\int_{0}^{\infty}\frac{dte^{-t/2}}{\sqrt{\cosh t-1}}
\left(1-\frac{t(1+e^{-t})}{1-e^{-2t}}\right).
\end{align}
\end{subequations}
These are convergent integrals, also found in Ref.~\cite{mazzitelli}, and their numerical evaluation gives $p\simeq -0.39$ and $q\simeq -1.41$. These approximate values for $p$ and $q$ can be readily verified by numerical integration.
As a consistency check, in the limit $\eta\to 0$ Eq.~\eqref{eq:phi2_final} reduces to Candelas' result for Schwarzschild~\cite{candelas}.

To make the decomposition in \eqref{eq:phi2_final} explicit, we first consider the purely monopole-induced contribution term~\cite{mazzitelli},
\begin{equation}
\label{eq:MLterm}
\left\langle \Psi^{2}\right\rangle_{\mathrm{ren}}^{\mathrm{(GM)}}
=
-\frac{\eta^{2}\!\left(p-2\xi q\right)}{8\sqrt{2}\pi^{2}r^{2}}
-\frac{\left(\xi-\frac{1}{6}\right)\eta^{2}}{8\pi^{2}r^{2}}\ln\!\left(\mu^{2}r^{2}\right),
\end{equation}
evaluated at the black hole horizon $r_{h}=2M/(1-\eta^{2})$ and expanded to $O(\eta^{2})$, which yields the first two terms in \eqref{eq:phi2_final}.

The remaining term must be compared with the pure Schwarzschild black hole, so we perform the rescaling \eqref{change of coordinates},
\begin{equation}
\bar{t}=\sqrt{1-\eta^2}\,t,
\qquad
\bar{r}=\frac{r}{\sqrt{1-\eta^2}},
\end{equation}
under which the Schwarzschild-global monopole metric \eqref{SMBH} becomes
\begin{equation}
\label{eq:metric_bar}
ds^{2}=-\left(1-\frac{2\bar{M}}{\bar{r}}\right)d\bar{t}^{2}+\frac{d\bar{r}^{2}}{\left(1-\frac{2\bar{M}}{\bar{r}}\right)}+\left(1-\eta^{2}\right)\bar{r}^{2}d\Omega^{2},
\end{equation}
where $\bar{M}=M/(1-\eta^{2})^{3/2}$. With this convention the horizon is located at
\begin{equation}
\bar{r}_{h}=2\bar{M}.
\end{equation} After this rescaling, the Schwarzschild contribution \cite{candelas} written in terms of the rescaled horizon radius becomes

\begin{equation}\label{eq:Psi2Schw}\begin{aligned}
\left\langle \Psi^{2}\right\rangle_{\mathrm{ren}}^{\mathrm{(Schw)}}
=\frac{1}{192\pi^{2}\bar{M}^{2}}
=\frac{(1-\eta^{2})^{3}}{192\pi^{2}M^{2}}
\approx\frac{1-3\eta^{2}}{192\pi^{2}M^{2}},
\end{aligned}\end{equation}
in agreement with the last term of \eqref{eq:phi2_final}.

Finally, we emphasize that, when evaluated at the event horizon, the purely monopole-induced contribution to 
$\langle\Psi^{2}\rangle_{\mathrm{ren}}$--namely, the first two terms in 
Eq.~\eqref{eq:phi2_final}--coincides with the result obtained for a \emph{naked} 
global monopole spacetime with Dirichlet boundary conditions imposed at $r=0$. 
In the Hartle-Hawking state, the Euclidean section is restricted to the exterior 
region, $r\in[r_h,\infty)$, and the corresponding Green function is uniquely fixed 
by regularity at the horizon together with the prescribed fall-off at infinity. 
This restriction removes the non-Dirichlet contributions that may arise in the 
naked global monopole case~\cite{pitelli2}. This point is essential: the additional 
non-Dirichlet terms discussed in Ref.~\cite{pitelli2} appear already at order 
$O(\eta^0)$ and would therefore obstruct the recovery of the Schwarzschild limit 
as $\eta\to0$. Hence, our result is consistent with Candelas' expression in the 
Schwarzschild limit. In addition, our result contrasts with Eq.(\ref{eq:Psi2Ottewill}) as it does not possesses any angular dependence. This is due to the globally symmetric character of the global monopole defect, resulting in an isotropic expression across the horizon.

A further consistency check on the additive decomposition obtained for $\langle \Psi^2\rangle_{\mathrm{ren}}$ is provided by the trace of the renormalized stress-energy tensor in the conformal case. For a massless, conformally coupled scalar field, the trace $\langle T^{\mu}_{\;\mu}\rangle_{\mathrm{ren}}$ is given by the conformal anomaly, which depends only on local geometric invariants and is independent of the quantum state. In four dimensions, the conformal anomaly reads \cite{Decanini:2005eg,Birrell:1982ix}:
\begin{equation}
    \langle T^{\mu}_{\;\mu}\rangle_{\mathrm{ren}}=\frac{1}{2880\pi^{2}}\left(R_{\mu\nu\rho\sigma}R^{\mu\nu\rho\sigma}-R_{\mu\nu}R^{\mu\nu}+\Box R\right).
\end{equation}

The curvature invariants for the Schwarzschild-global monopole metric are obtained by straightforward computation. They consist of the Kretschmann scalar $R_{\mu\nu\rho\sigma}R^{\mu\nu\rho\sigma}$, the Ricci‑squared term $R_{\mu\nu}R^{\mu\nu}$, and the d’Alembertian of the Ricci scalar $\Box R$. Explicitly, they read
\begin{align}
R_{\mu\nu\rho\sigma}R^{\mu\nu\rho\sigma}
&= \frac{48M^{2}}{r^{6}}+\frac{16M\eta^{2}}{r^{5}}+\frac{4\eta^{4}}{r^{4}}, \\
R_{\mu\nu}R^{\mu\nu}
&= \frac{2\eta^{4}}{r^{4}}, \\
\Box R
&= \frac{4\eta^{2}\left(1-\eta^{2}\right)}{r^{4}}-\frac{16M\eta^{2}}{r^{5}}.
\end{align}

Notice that both $R_{\mu\nu\rho\sigma}R^{\mu\nu\rho\sigma}$ and $\Box R$ contain terms linear in $M\eta^{2}/r^{5}$. However, when these expressions are inserted into the anomaly, these mixed terms cancel each other. Substituting the above, we obtain the general result
\begin{equation}
   \left\langle T_{\mu}^{\mu}\right\rangle _{\text{ren}}=\frac{\eta^{2}}{720\pi^{2}r^{4}}\left(1-\frac{\eta^{2}}{2}\right)+\frac{M^{2}}{60\pi^{2}r^{6}}.
\end{equation}
It is worth noting that this expression consists of the individual contributions from the global monopole and the Schwarzschild black hole, presented in that order. Evaluating it on the event horizon $r_h = 2M/(1-\eta^2)$ yields
\begin{equation}
   \left\langle T_{\mu}^{\mu}\right\rangle _{\text{ren}}=\frac{\eta^{2}\left(1-\eta^{2}\right)^{4}}{11520\pi^{2}M^{4}}\left(1-\frac{\eta^{2}}{2}\right)+\frac{\left(1-\eta^{2}\right)^{6}}{3840\pi^{2}M^{4}}.
\end{equation}

We now expand this expression in powers of the small monopole parameter $\eta$, since we are working perturbatively to $\mathcal{O}(\eta^2)$. Expanding the factors $(1-\eta^2)^6$ and $(1-\eta^2)^4$ to linear order in $\eta^2$, and multiplying by the prefactors, we find after simplification
\begin{equation}
   \left\langle T_{\mu}^{\mu}\right\rangle _{\text{ren}}=\frac{\eta^{2}}{11520\pi^{2}M^{4}}+\frac{1-6\eta^{2}}{3840\pi^{2}M^{4}}+\mathcal{O}\left(\eta^{4}\right).
\end{equation}

The first term is precisely the conformal anomaly of a naked global monopole evaluated at the horizon radius $r_h$. The second term coincides with the Schwarzschild conformal anomaly under the rescaling $M\rightarrow M/(1-\eta^2)^{3/2}$, the same rescaling that appears for $\langle \Psi^{2}\rangle_{\mathrm{ren}}$ in Eq.(\ref{eq:Psi2Schw}). Thus, the additive decomposition uncovered for $\langle \Psi^{2}\rangle_{\mathrm{ren}}$ is mirrored in the conformal anomaly, which is a purely geometric quantity.

\section{Conclusions}\label{Discussion_and_conclusions}

In this work, we have investigated the renormalized vacuum polarization, $\langle\Psi^{2}\rangle_{\text{ren}}$, for a massless scalar field with arbitrary curvature coupling $\xi$ in the background of a Schwarzschild black hole carrying a global monopole. By utilizing a Green’s function approach within the Hartle-Hawking state and performing a Wick rotation to the Euclidean section, we derived the near-horizon behavior of vacuum fluctuations. Our analysis was conducted perturbatively in the monopole parameter $\eta$, retaining terms up to $\mathcal{O}(\eta^2)$ to capture the leading-order corrections to the black hole geometry.

The central result of our study is the additive decomposition of the vacuum polarization on the event horizon. We found that the renormalized expectation value splits into two independent parts: (i) a monopole-induced contribution, which matches the results for a pure global monopole spacetime when evaluated at the event horizon radius $r_h$, and (ii) the standard Schwarzschild contribution, originally calculated by Candelas, expressed in terms of a modified horizon radius that accounts for the presence of the global monopole.

As an independent consistency check, we have shown that the conformal anomaly, a purely geometric quantity, exhibits the same additive decomposition when evaluated on the horizon. The monopole part reproduces the conformal anomaly of a naked global monopole, while the Schwarzschild part emerges under the same mass rescaling that appears for $\langle \Psi^2 \rangle_{\mathrm{ren}}$. This reinforces the structural result obtained from the vacuum polarization and offers preliminary insights required for the subsequent derivation of the full $\langle T_{\mu\nu} \rangle_{\text{ren}}$.

This structural result parallels earlier findings for Schwarzschild black holes pierced by a cosmic string, reinforcing the idea that vacuum polarization on the horizon responds to topological defects in a predictably fashion. While naked global monopole spacetimes require an explicit choice of boundary conditions at the origin, we demonstrated that in the black hole case, the Green's function is uniquely determined by regularity at the event horizon and the behavior at infinity. The consistency of our results with the pure Schwarzschild limit as $\eta \to 0$ confirms that the presence of the horizon naturally selects the physical branch of the solution.

Ultimately, this study highlights how the interplay between the solid-angle deficit of a global monopole and the event horizon of a black hole manifests in quantum observables. Given that $\langle\Psi^{2}\rangle_{\text{ren}}$ serves as a technically simpler proxy for the renormalized stress-energy tensor, these results provide insights toward a full evaluation of $\langle T_{\mu\nu} \rangle_{\text{ren}}$ in this spacetime.

\section{Acknowledgments}\label{Acknowledgements}
L. G. B. would like to thank the Coordenação de Aperfeiçoamento de Pessoal de Nível Superior--Brazil (CAPES)--Finance Code 001 for the financial support. V. H. M. R. acknowledges the financial support of Coordenação de Aperfeiçoamento de Pessoal de Nível Superior (CAPES)--Brazil, Finance Code 001, and gratefully acknowledges the kind hospitality of the Department of Mathematics of the University of Genova. J. P. M. P.
thanks the support provided in part by Conselho Nacional de Desenvolvimento Cient\'ifico e Tecnol\'ogico (CNPq, Brazil), Grant No. 305194/2025-9, and Funda\c{c}\~ao de Amparo a Pesquisa do Estado de S\~ao Paulo (FAPESP), Grant No. 2024/00923-6.

\bibliography{sample}

\end{document}